\documentclass[letterpaper,10pt]{article}
\usepackage{cite,bm,graphicx}

\graphicspath{{images/}}

\begin{document}

\date{\today}

\title{Modeling the isotropic/smectic-C tilted lamellar liquid crystalline transition}
 
\author{
Nasser Mohieddin Abukhdeir and Alejandro D. Rey\\
Department of Chemical Engineering\\
McGill University\\
Montreal, QC, Canada\\
nasser.abukhdeir@mcgill.ca \\
alejandro.rey@mcgill.ca\\
}

\date{\today}
\maketitle

\begin{abstract}
Extensions of a previously presented Landau-de Gennes type liquid crystalline phase transition model for the direct isotropic/smectic-A (lamellar) a phase transition to the direct isotropic/smectic-C (tilted lamellar) transition are studied.  Two different proposed extensions to the model are studied both in the context of ideal ordering (point volume) and full three-dimensional uniaxial scalar/vector decomposition.  Recommendations based upon the inclusion of essential physics and computational feasibility are made, distinguishing each of the two proposed extensions based upon these criteria.  Additionally, it is found that the approach used for the isotropic/smectic-A model to deterministically compute phase diagram data is not possible for either of the two proposed isotropic/smectic-C models.
\end{abstract}

\noindent
{\bf\normalsize KEY WORDS}\newline
{liquid crystals smectic-c modeling simulation lamellar tilt}

\section{Introduction}
Liquid crystalline phase ordering or self-organization is a pervasive phenomena observed throughout the physical world.  The study of this class of soft matter is becoming increasingly more prevalent due to liquid crystal-based technological advances in the past few decades.  Apart from the development of new types of display technology and high-performance materials, much interest is focused on mesophases (liquid crystalline phases) in Nature.  In order to accomplish this task, fundemental understanding of mesophases must be extended beyond the simplest types of liquid crystals, nematics, where some degree of orientational order is present.  Higher order smectic and columnar mesophases, where additionally there is some degree of translational order, have been found in a vast number of biological materials and processes.  Harnessing the self-organizing properties of liquid crystals to solve engineering problems can thus be extended from the current state of the art, manipulation of bulk material properties, to more complex techniques such as biosensing and biomimetics .

One of the main challenges inhibiting the study of smectic liquid crystals are the time ($ns$) and length ($nm$) scales of their molecular organization.  As a result, experimental approaches are mainly limited to capturing static phenomena.  However, theoretical approaches have been able to both enhance experimental techniques and independently uncover unknown physics.  Numerical simulation, harnessing recent computational advances, are at the forefront of these theoretical approaches.  In this context, simulations based on continuum models are the most viable approach to studying multiple scales, both experimentally accessible and inaccessible.

The goal of past and ongoing work studying the most simple of the smectic phases, the lamellar smectic-A mesophase, has shown great promise \cite{Abukhdeir2008a,Abukhdeir2008c}.  Applying similar approaches to the study of a more complex smectic, the tilted lamellar smectic-C mesophase, is therefore a clear progression.  A first step to modeling and simulation of these and other materials is the development of a suitable model and the incorporation of existing experimental data into that model, when possible.  To achieve this goal the problem can be approached through the development of a continuum model that has characteristics conducive to representing the observed physical system.  In this work, two different extensions of a Landau-de Gennes type model mesophase transition model \cite{deGennes1995,Mukherjee2002a,Mukherjee2005,Biscari2007} will be analyzed from this viewpoint.  The objective of this work is to determine the suitability of the two models for use in both phase diagram computation and full three-dimensional simulation of the isotropic/smectic-C mesophase transition.

The paper is organized in this way: first a brief introduction will address relevent liquid crystal mesophases and the theortical approach used to characterize them.  The Landau-de Gennes type phase transition model for the direct isotropic/smectic-A model will then be presented and features explained in the context of these types of theoretical approaches.  The extensions of this model to the isotropic/smectic-C transition will then be presented and discussed.  Finally, conclusions regarding the suitability of each extension will be made based upon both the extent to which each model incorporates essential physics and feasibility of simulation.

\section{Background} \label{sec:background}
\subsection{Liquid Crystal Mesophases} \label{sec:mesophases}
Liquid crystals or mesophases are partially ordered materials that flow like liquids but have some degree of orientational/translation order like crystals.  They are composed of anisotropic molecules called mesogens which can be discotic (disc-like) or calamatic (rod-like).  Thermotropic liquid crystals are typically pure-component compounds that exhibit mesophase ordering most greatly in response to temperature changes.  Lyotropic liquid crystals are mixtures of mesogens, possibly with a solvent, that most greatly exhibit mesophase behavior in response to concentration changes.  Effects of pressure and flow fields/bulk effects also influence mesophase behavior.  This work will focus on temperature effects on thermotropic calamatic liquid crystals.

\begin{figure}[htp] 
\includegraphics[width=5in]{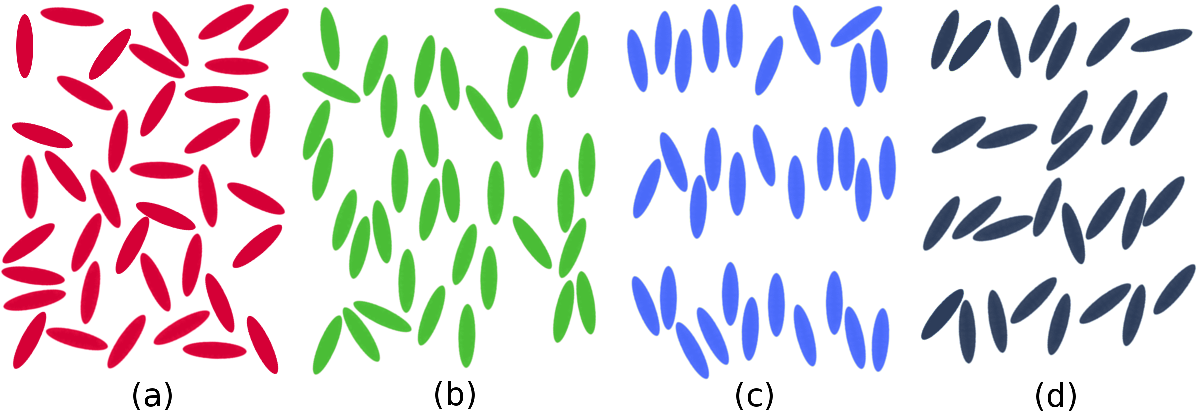} 
\caption[Graphical Mesophase Ordering]{Graphical representation of mesophase ordering: a) isotropic b) nematic c) smectic-A and d) smectic-C \label{fig:mesophase}}
\end{figure}

An unordered liquid where, there is neither orientational nor translation order of the molecules, is referred to as isotropic.  Of the many different mesophases observed, the three of interest in this work are the nematic, smectic-A, and smectic-C.  These mesophases have increasing degrees of liquid crystalline ordering.  The nematic phase exhibits purely orientational order where molecules conform to an average molecular axis, the nematic director.  Smectic-A ordering has orientational order as with nematics but, in addition, has one-dimensional translational order in the direction of the average molecular axis.  This can be thought of in terms of two-dimension fluid layers stacked upon each other.  Finally, the smectic-C mesophase is identical to the smectic-A, except the average molecular axis forms an angle with the normal to the layers.  Schematic representations of these four phases are shown in Figure \ref{fig:mesophase}.
Theoretical characterization of mesophase order is accomplished using order parameters that adequately capture the physics involved.  These order parameters typically have an amplitude and phase associated with them.  In order to characterize the partial orientational order of the nematic phase, a second order symmetric traceless tensor can be used \cite{deGennes1995}:
\begin{equation} \label{eqnem_order_param}
\bm{Q} = S \left(\bm{nn} - \frac{1}{3} \bm{I}\right) + \frac{1}{3} P \left( \bm{mm} - \bm{ll}\right)
\end{equation}
where $\mathbf{n}/\mathbf{m}/\mathbf{l}$ are the eigenvectors of $\bm{Q}$, which characterize the average molecular orientational axes, and $S/P$ are scalars which represent the extent to which the molecules conform to the average orientational axes \cite{Rey2002,Yan2002,Rey2007}.  Uniaxial order is characterized by $S$ and $\bm{n}$, which correspond to the maximum eigenvalue(and its corresponding eigenvector) of the Q-tensor, $S= \frac{3}{2} \mu_n$.  Biaxial order is characterized by $P$ and $\bm{m}/\bm{l}$, which correspond to the lesser eigenvalues and eigenvectors, $P = -\frac{3}{2}\left(\mu_m - \mu_l\right)$.

The smectic-A and smectic-C mesophases have one-dimensional translational order in addition to the orientational order found in nematics.  Characterizing these mesophases can be accomplished through the use of primary (orientational) and secondary (translational) order parameters together \cite{Toledano1987}.  This is accomplished using the tensor order parameter (\ref{eqnem_order_param}) and the complex order parameter \cite{deGennes1995}:
\begin{equation} \label{eqsmec_order_param}
\Psi = \psi e^{i \phi}
\end{equation}
where $\phi$ is the phase, $\psi$ is the scalar amplitude of the density modulation.  The density wave vector, which describes the average orientation of the smectic-A/smectic-C density modulation, is defined as $\mathbf{a} = \nabla \phi / |{\nabla \phi}|$.  The smectic scalar order parameter $\psi$ characterizes the magnitude of the density modulation, and is used in a dimensionless form in this work.  In the smectic-A mesophase the nematic director and the density wave vector prefer parallel orientation.  In the smectic-C mesophase a tilt angle $\theta$ is observed between the nematic director and smectic wave vector.

The isotropic/smectic mesophase transition is first-order, where the materials exhibit discontinuities in their physical properties across the phase transition.  These transitions have a latent heat associated with them and a regime in which both the disordered and ordered phases coexist in a metastable state.  This phase transition mechanism involves nucleation and growth.  Second order transitions are have a continuous evolution of the material properties and the growth mechanism is through spinodal decomposition.

\subsection{Landau-de Gennes type model for the direct isotropic/ smectic-A transition}
A Landau-type free energy expansion is a Taylor series expansion of the free energy of the system.  The terms included in this expansion are have a great effect on both the phenomena that can be captured and the computational difficulty in applying the model.  These two competing goals along with the use of multiple non-scalar order parameters complicates the matter.  In this particular model three types of terms can be distinguished from each other:  homogeneous, gradient, and coupling terms.  Only terms up to the fourth order are included \cite{Mukherjee2001} and in addition to this it is assumed that in the vicinity of the transition only the coefficients of the second order terms vary significantly with temperature \cite{Singh2002}.  

A Landau-de Gennes type model for the first order isotropic/smectic-A phase transition is used based upon a model initially presented by Mukherjee, Pleiner, and Brand \cite{deGennes1995,Mukherjee2001}:
\begin{eqnarray} \label{eq:free_energy_heterogeneous}
f_S - f_0 &=&\frac{1}{2} a \left(\bm{Q} : \bm{Q}\right) - \frac{1}{3} b \left(\bm{Q}\cdot\bm{Q}\right) : \bm{Q} + \frac{1}{4} c \left(\bm{Q} : \bm{Q}\right)^2 + \frac{1}{2} \alpha \left|\Psi\right|^2 + \frac{1}{4} \beta \left|\Psi\right|^4 \nonumber\\
&&- \frac{1}{2} \delta \psi^2 \left(\bm{Q} : \bm{Q}\right) - \frac{1}{2} e \bm{Q}:\left(\bm{\nabla} \Psi\right)\left(\bm{\nabla} \Psi^*\right) \nonumber\\
&& + \frac{1}{2} l_1 \left(\bm{\nabla} \bm{Q} \right)^2 + \frac{1}{2} l_2 \left(\bm{\nabla} \cdot \bm{Q} \right)^2 + \frac{1}{2} l_3 \bm{Q}: \left(\bm{\nabla}\bm{Q} : \bm{\nabla}\bm{Q}\right) \nonumber\\
&& + \frac{1}{2} b_1 \left|\bm{\nabla} \Psi\right|^2 + \frac{1}{4} b_2 \left|\nabla^2 \Psi\right|^2
\end{eqnarray}

\begin{eqnarray} \label{eq:free_energy_heterogenous_coeffs}
A & = & a_0 (T - T_{NI}) \nonumber \\
\alpha & = & \alpha_0 (T - T_{AI})\nonumber 
\end{eqnarray}
where $f$ is the free energy density, $f_0$ is the free energy density of the isotropic phase, terms 1-5 are the bulk contributions to the free energy, terms 6-7 are couplings of nematic and smectic order; both the bulk order and coupling of the nematic director and smectic density-wave vector, respectively.  Terms 8-10 are the nematic and terms 11-12 smectic elastic contributions to the free energy, respectively.  The order parameters are defined in (\ref{eqnem_order_param}-\ref{eqsmec_order_param}), $T$ is temperature, $T_{NI}$/$T_{AI}$ are the hypothetical second order transition temperatures for isotropic/nematic and isotropic/smectic-A mesophase transitions, and the remaining constants are phenomenological parameters.  Further explanation and justification for the use of this high-order model can be found in \cite{Abukhdeir2008a}.

The first extension to the isotropic/smectic-A model eqn. \ref{eq:free_energy_heterogeneous} to the isotropic/smectic-C model was proposed by Mukherjee, Pleiner, and Brand \cite{Mukherjee2002a} and then later modified \cite{Mukherjee2005}:

\begin{equation} \label{eq:free_energy_smec_c1}
f_C -f_0 = f_S + \frac{1}{2} f \left( \bm{Q} \cdot \bm{Q}\right):\bm{\nabla} \Psi \bm{\nabla} \Psi^* + \frac{1}{4} h \left( \bm{Q} : \bm{\nabla} \bm{\nabla} \Psi \right) \left( \bm{Q} : \bm{\nabla} \bm{\nabla} \Psi^* \right)
\end{equation}

Later, Biscari,Calderer, and Terentjev \cite{Biscari2007} proposed an alternate extension to eqn. \ref{eq:free_energy_heterogeneous}:

\begin{equation} \label{eq:free_energy_smec_c2}
f_C -f_0 = f_S + \frac{1}{4} f \left( \bm{Q} :\bm{\nabla} \Psi \bm{\nabla} \Psi^* \right)^2
\end{equation}

\section{Results and Discussion}

\subsection{Homogeneous volume comparison}

Eqns. \ref{eq:free_energy_smec_c1} and \ref{eq:free_energy_smec_c2} can be simplified for use in phase diagram computation by the assumption of a homogeneous volume.  This homogeneous free energy equation  is a function of four scalar components of the two original order parameters related to the degree of orientational order, positional order, the magnitude of the density wave, and the tilt angle formed between the density wave and the nematic director.  Thus, following ref. \cite{Mukherjee2002a}, the smectic density wave is fixed along the z-axis and the nematic director $\mathbf{n}$ takes the form:
\begin{equation}
\bm{n} = cos(\theta) \bm{z} + sin(\theta)\bm{y}
\end{equation}
where the unit vector $\bm{z}$ is chosen parallel to the wave vector and $\bm{y}$ is an arbritary orthogonal vector to $\bm{z}$.  Note that gradients of $\Psi$ does not vanish due to the inherent gradient of the density wave.  The simplified free energy equation for the extension proposed by Mukherjee et al \cite{Mukherjee2002a,Mukherjee2005} is:
\begin{eqnarray} \label{eq:free_energy_homogeneous1}
f - f_0  &=&  \frac{1}{3} a S^2 - \frac{2}{27} b S^3+ \frac{1}{9} c S^4 + \frac{1}{2} \alpha \psi^2 + \frac{1}{4} \beta \psi^4 \nonumber\\
&& - \frac{1}{3} \delta  S^2 \psi^2 - \frac{1}{2} e S \psi^2 q^2 \left(\cos^2{\theta}- \frac{1}{3}\right) + \frac{1}{2} f S^2 \psi^2 q^2 \left(\cos^2{\theta} + \frac{1}{3}\right)  \nonumber\\
&&+ \frac{1}{4} h S^2 \psi^2 q^4 \left(\cos^2{\theta} - \frac{1}{3}\right)^2 \nonumber\\
&&+ \frac{1}{2} b_1 \psi^2 q^2 + \frac{1}{4} b_2 \psi^2 q^4 
\end{eqnarray}
where $q=2 \pi/d$ is the magnitude of the wave vector and $d$ is the smectic layer spacing.  The minima criteria for the tilt angle $\theta$ involves the partial derivative of the homogeneous free energy with respect to $\theta$:
\begin{equation} \label{eq:homo_smec_c1}
\frac{\partial f_C}{\partial \theta} = 0 = e S \psi^2 q^2 \cos{\theta}\sin{\theta} - f S^2 \psi^2 q^2 \cos{\theta}\sin{\theta} - h S^2 \psi^2 q^4 \left(\cos^2{\theta} - \frac{1}{3}\right) \cos{\theta}\sin{\theta}
\end{equation}
Assuming the smectic-C phase is present $S,\psi,q,\theta>0$ and eqn. \ref{eq:homo_smec_c1} simplifies to:
\begin{equation} 
0 = e - f S  - h S q^2 \left(\cos^2{\theta} - \frac{1}{3}\right)
\end{equation}
Clearly the competition of the $e/f$-terms with the $h$-term can result in a finite value of the tilt angle $\theta$, but the addition of the $f$ term is not higher order in $\theta$ (compared to the $e$-term), thus can be neglected if it does incorporate necessary physics (determined in the next section). 
The simplified free energy equation for the extension proposed by Biscari et al \cite{Biscari2007} is: 
\begin{eqnarray} \label{eq:free_energy_homogeneous2}
f - f_0  &=&  \frac{1}{3} a S^2 - \frac{2}{27} b S^3+ \frac{1}{9} c S^4 + \frac{1}{2} \alpha \psi^2 + \frac{1}{4} \beta \psi^4 \nonumber\\
&& - \frac{1}{3} \delta  S^2 \psi^2 - \frac{1}{2} e S \psi^2 q^2 \left(\cos^2{\theta}- \frac{1}{3}\right) + \frac{1}{2} f S^2 \psi^4 q^4 \left(\cos^2{\theta} - \frac{1}{3}\right)^2  \nonumber\\
&&+ \frac{1}{2} b_1 \psi^2 q^2 + \frac{1}{4} b_2 \psi^2 q^4 
\end{eqnarray}
which results in a minima criteria for the tilt angle $\theta$:
\begin{equation} \label{eq:homo_smec_c2}
\frac{\partial f_C}{\partial \theta} = 0 = e S \psi^2 q^2 \cos{\theta}\sin{\theta} - f S^2 \psi^4 q^4 \left(\cos^2{\theta} - \frac{1}{3}\right) \cos{\theta}\sin{\theta} 
\end{equation}
Assuming the smectic-C phase is present, eqn. \ref{eq:homo_smec_c2} can be simplified to:
\begin{equation} 
 0 = e  - f S \psi^2 q^2 \left(\cos^2{\theta} - \frac{1}{3}\right) 
\end{equation}
The competition of the $e$- and $f$-terms in this model also results in a finite value of the tilt angle $\theta$.

\subsection{Scalar/vector decomposition}

Both models can now be decomposed into scalar/vector form, using the assumption of uniaxial orientation such that the order parameters are $S$ the uniaxial nematic order parameter, $\bm{n}$ the nematic director, $\psi$ the smectic scalar order parameter, and $\bm{a} = \nabla \phi$ the smectic wave vector.  Focusing on the nematic/smectic coupling terms $e$, $f$, and $h$, the model of Mukherjee et al \cite{Mukherjee2002a,Mukherjee2005} results in:
\begin{eqnarray}
- \frac{1}{2} e S \left[ \bm{nn}:\left( \nabla\psi \nabla\psi + \bm{aa}\right) - \frac{1}{3} \left( (\nabla \psi)^2 + \psi^2 (\bm{a}^2)\right)\right] \nonumber\\
+ \frac{1}{6} f S^2 \left[ \bm{nn}:\left(  \nabla\psi \nabla\psi + \bm{aa}\right) + \frac{1}{3}\left( (\nabla \psi)^2 + \psi^2 (\bm{a}^2)\right)\right]\nonumber\\
+ \frac{1}{4}\left[ \bm{nnnn}::\bm{\Gamma} + \frac{1}{9}\bm{\delta\delta}::\bm{\Gamma} - \frac{1}{3}\left(\bm{nn\delta}+\bm{\delta nn}\right)::\bm{\Gamma}\right]\\
\bm{\Gamma} = \nabla\nabla \psi \nabla\nabla \psi + 4 (\bm{a} \nabla \psi)  (\bm{a}\nabla \psi )+ \psi^2 \left(\nabla\nabla \phi \nabla\nabla \phi\right)
\end{eqnarray}
clearly both the $e$- and $f$-terms are first-order projections of the nematic director onto the smectic wave vector.  The inclusion of the $e$ term includes the essential physics of the first-order projection, and thus the $f$ term can be neglected.  The higher-order $h$ term is necessary in order to have a finite tilt angle, but the physical significance of the higher-order director/wave-vector projection is not clearly essential to the model.  This term includes coupling of the nematic director to the layer curvature, in addition to accounting for both average and Gaussian curvature of the smectic layers.  Average curvature of the layers is accounted for in the existing $b_2$ term (eqn. \ref{eq:free_energy_heterogeneous}) and the inclusion of Gaussian curvature would have a negligible on model predictions.  Additionally, the phenomenological coefficient itself would be infeasible to measure experimentally or to determine through molecular simulation. 
The model of Biscari et al \cite{Biscari2007} results in a simpler set of nematic/smectic coupling terms:
\begin{eqnarray}
- \frac{1}{2} e S \left[ \bm{nn}:\left( \nabla\psi \nabla\psi + \bm{aa}\right) - \frac{1}{3} \left( (\nabla \psi)^2 + \psi^2 (\bm{a}^2)\right)\right] \nonumber\\
+ \frac{1}{4} f S^2 \left[ \bm{nn}:\left(  \nabla\psi \nabla\psi + \bm{aa}\right) + \frac{1}{3}\left( (\nabla \psi)^2 + \psi^2 (\bm{a}^2)\right)\right]^2
\end{eqnarray}
where now the $f$-term is simply second-order in the $e$-term.  This simple approach both allows for a finite tilt angle and ease of numerical simulation.

\section{Conclusion \label{sec:conc}}
A theoretical study of two previously proposed models for the direct isotropic/smectic-C phase transition \cite{Mukherjee2002a,Mukherjee2005,Biscari2007} were studied in the context of ideal ordering (point volume) and full three-dimensional uniaxial scalar/vector decomposition.  Based upon the inclusion of essential physics and computational feasibility, the model of Biscari et al \cite{Biscari2007} was found to be the most computationally feasible of the two models.  Nonetheless, past approaches to deterministically compute phase diagrams for these types of models \cite{Abukhdeir2007,Abukhdeir2008c} would not be possible due to the additional couplings introduced through the tilt angle.

\section{Acknowledgements} \label{sec:ack}
This research was partially supported by the Natural Sciences and Engineering Council of Canada.

\bibliographystyle{unsrt}
\bibliography{/home/nasser/nfs/references/references}

\begin{thebibliography}{10}

\bibitem{Abukhdeir2008a}
N.M Abukhdeir and A.D Rey.
\newblock Defect kinetics and dynamics of pattern coarsening in a
  two-dimensional smectic-a system.
\newblock {\em New Journal of Physics}, 10(6):063025 (17pp), 2008.

\bibitem{Abukhdeir2008c}
N.M. Abukhdeir and A.D. Rey.
\newblock Simulation of spherulite growth using a comprehensive approach to
  modeling the first-order isotropic/smectic-a mesophase transition.
\newblock {\em Arxiv preprint arXiv:0807.4525}, 2008.
\newblock Submitted to Communications in Computational Physics July 2008,
  manuscript ID CFluids08-01.

\bibitem{deGennes1995}
P.G. de~Gennes and J~Prost.
\newblock {\em The Physics of Liquid Crystals}.
\newblock Oxford University Press, New York, second edition, 1995.

\bibitem{Mukherjee2002a}
Prabir~K. Mukherjee, Harald Pleiner, and Helmut~R. Brand.
\newblock Landau model of the smectic c--isotropic phase transition.
\newblock {\em The Journal of Chemical Physics}, 117(16):7788--7792, 2002.

\bibitem{Mukherjee2005}
P.~K. Mukherjee, H.~Pleiner, and H.~R. Brand.
\newblock A phenomenological theory of the isotropic to chiral smectic-c phase
  transition.
\newblock {\em European Physical Journal E: Soft Matter}, 17:501--506, 2005.

\bibitem{Biscari2007}
P.~Biscari, M.C. Calderer, and E.M. Terentjev.
\newblock Landau de gennes theory of isotropic-nematic-smectic liquid crystal
  transitions.
\newblock {\em Phys Rev E Stat Nonlin Soft Matter Phys}, 75(5):051707, May
  2007.

\bibitem{Rey2002}
AD~Rey and M~Denn.
\newblock Dynamical phenomena in liquid-crystalline materials.
\newblock {\em Annual Review of Fluid Mechanics}, 34(1):p233 --, 2002.

\bibitem{Yan2002}
J.~Yan and A.~D. Rey.
\newblock Texture formation in carbonaceous mesophase fibers.
\newblock {\em Phys. Rev. E}, 65(3):031713, Feb 2002.

\bibitem{Rey2007}
Alejandro~D. Rey.
\newblock Capillary models for liquid crystal fibers, membranes, films, and
  drops.
\newblock {\em Soft Matter}, 3:1349 -- 1368, 2007.

\bibitem{Toledano1987}
Jean-Claude Toledano and Pierre Toledano.
\newblock {\em The Landau Theory of Phase Transitions: Application to
  Structural, Incommensurate, Magnetic, and Liquid Crystal Systems (World
  Scientific Lecture Notes in Physics)}.
\newblock World Scientific Pub Co Inc, 1987.

\bibitem{Mukherjee2001}
P.~K. Mukherjee, H.~Pleiner, and H.~R. Brand.
\newblock Simple landau model of the smectic-a-isotropic phase transition.
\newblock {\em European Physical Journal E: Soft Matter}, 4:293--297, 2001.

\bibitem{Singh2002}
Shri Singh.
\newblock {\em Liquid Crystals: Fundamentals}.
\newblock World Scientific Publishing Company, 1st edition, 7 2002.

\bibitem{Abukhdeir2007}
N.M. Abukhdeir and A.D. Rey.
\newblock Nonlinear model for the isotropic/smectic a phase transition.
\newblock In {\em Modelling and Simulation}. International Association of
  Science and Technology for Development (IASTED), ACTA Press, 2007.

\end{thebibliography}

\end{document}